\documentclass{article}
\usepackage[labelformat=simple]{subcaption}

\usepackage[english]{babel}

\usepackage[letterpaper,top=2cm,bottom=2cm,left=3cm,right=3cm,marginparwidth=1.75cm]{geometry}

\usepackage{amsmath}
\usepackage{lscape}
\usepackage{graphicx}
\usepackage[colorlinks=true, allcolors=blue]{hyperref}
\usepackage{mathptmx, amsthm, mathtools, mathrsfs, amssymb, ragged2e}

\title{\textbf{Evolution of Virial Clouds - II: From the Formation of First Stars up to their Explosion}}

\author{Noraiz Tahir $^{1,*}$, Asghar Qadir $^{2, **}$, Francesco De Paolis $^{3, 4, 5, \dagger}$, Noor Fatima$^{1, \ddagger}$ \\
$^1$ Department of Physics, School of Natural Sciences (SNS), National University of \\ Sciences and Technology (NUST), Sector H--12, 44000, Islamabad, Pakistan. \\
$^2$ Pakistan Academy of Sciences, Sector G--5/2, Islamabad, Pakistan. \\ 
$^3$ Department of Mathematics and Physics ``Ennio De Giorgi'', University of Salento, \\ Via per Arnesano, 73100, Lecce, Italy. \\
$^4$ INFN, Sezione di Lecce, Via per Arnesano, 73100, Lecce, Italy.\\
$^5$ INAF, Sezione di Lecce, Via per Arnesano, 73100, Lecce, Italy\\
$^*$ \textbf{corresponding author:} noraiztahir78637@gmail.com \\
$^{**}$ asgharqadir46@gmail.com \\
$^\dagger$ francesco.depaolis@le.infn.it\\
$^\ddagger$ noorfatima.nfs.offical@gmail.com\\
}

\begin{document}
\maketitle

\begin{abstract}
 The existence of cold gas and dust clouds close to the cosmic microwave background (CMB) temperature was proposed as a potential repository for a significant fraction of the missing baryons in galactic halos. While the evolution of the virial clouds from the last scattering surface (LSS) ($z=1100$) up to the formation of Population III (Pop III) stars ($z=48$) was studied in [N. Tahir, A. Qadir, M. Sakhi, \& F. De Paolis, The Euro. Phys. Jour. C 81, 827 (2021)] (Paper I), the subsequent evolution during the epoch of the first stars remained unexplored. In the present work, we investigate the second evolutionary phase of the virial clouds, covering the redshift range $48 \gtrsim z \gtrsim 10$. We develop a comprehensive model incorporating the coupled thermal, chemical, and dynamical evolution of the cloud, including CMB heating, adiabatic compression due to gravitational contraction, H$_2$ and metal-line cooling, and feedback from Pop III supernovae. Solving the coupled ordinary differential equations (ODE) numerically, we find that the cloud cools monotonically from $137$~K to $41.5$~K, contracts from $120$~pc to $89.8$~pc, and becomes enriched in metals and molecular hydrogen, while the cooling rate dominates the heating rate throughout the evolution.
\end{abstract}


\section{Introduction}

The nature and distribution of energy-matter in the Universe remains one of the outstanding puzzles in modern cosmology \cite{planck2018results,peebles2003dark,bertone2005dark,freese2017dark}. A significant fraction of the baryons predicted by Big Bang nucleosynthesis remains unaccounted for in the local Universe, a problem known as the ``missing baryon problem'' \cite{mcgaugh2007}. While the Lyman-$\alpha$ forest traces virtually all baryons at high redshifts, it accounts for only $\sim 30\%$ at low redshifts \cite{shull2012baryon}. Hydrodynamical simulations suggest that a substantial portion, perhaps $30$--$40\%$, resides in the warm-hot intergalactic medium (WHIM) \cite{shull2012baryon,nicastro2018missing}, though its direct detection remains challenging due to its low density and high ionization state \cite{nicastro2018missing,eckert2015whim}. Even with these advancements, a non-negligible fraction of the baryonic budget remains unaccounted for, motivating the search for alternative repositories.

A compelling proposal is that a significant portion of the missing baryons resides in galactic halos as cold gas clouds, which are difficult to detect through conventional emission or absorption techniques \cite{depaolis1995chameleon,qadir2019virial}. These clouds could contribute to the dark matter halo mass budget without violating observational constraints \cite{de1995galactic,qadir2019virial}. Their presence would naturally explain the observed frequency-independent temperature asymmetries in the CMB towards nearby galaxies, which are interpreted as Doppler shifts induced by the rotation of these clouds in galactic halos \cite{2011WMAPdepaolis,2014planckdepaolis,2015planckdepaolis,2015planckgurzadyan,2016planckdepaolis,2018planckgurzadyan,2019planckdepaolis}. The detected asymmetry has been used to trace galactic halo rotation \cite{tahir2019constraining,tahir2019ajom,tahir2023symmetry}, with contributions from the rotational kinetic Sunyaev-Zeldovich (rkSZ) effect and anomalous microwave emission (AME) from dust grain found to be subdominant \cite{2025rkszdepaolis,2025hvcstahir,2024rksztahir,2025amedepaolis}. These results provide indirect evidence for the presence and significance of virial clouds in galactic halos.

The existence of virial clouds was not immediately obvious theoretically. It was argued that equilibrium with the CMB at its current low temperature would be impossible because no internal modes could be excited by low-energy photons \cite{padmanabhan1990cmb}. However, Ref. \cite{qadir2019virial} demonstrated that equilibrium can arise through the translational mode of the gas; despite the extremely small probability of photon-molecule interactions, the large physical sizes and long thermalization timescales ensure that thermalization occurs faster than collapse. These clouds are thus pressure-supported by the CMB and remain in quasi-static equilibrium with the heat bath.

To test the virial cloud proposal against observations, it is necessary to trace their full evolutionary history from the LSS to the present day. The first stage, from $1100 \gtrsim z \gtrsim 48$, was studied in Paper~I \cite{tahir2021evolutionI}, where the clouds, with primordial composition ($X_{\rm H}\sim 0.75$, $X_{\rm He}\sim 0.25$, $x_{\rm H_2}\sim 10^{-6}$), remained in quasi-static equilibrium with the adiabatically cooling CMB, becoming denser, smaller, and less massive, with the Jeans mass decreasing from $\sim 10^5\,M_\odot$ at $z=1100$ to $\sim 10^4\,M_\odot$ at $z=48$ \cite{peebles1968h2,lepp1984h2}.

The present work addresses the second phase, spanning $48 \gtrsim z \gtrsim 10$, during which the physical environment is fundamentally altered by the formation of Pop~III stars. These massive stars $\sim 10^2\,M_\odot$ live for $\sim 10^6$~yr and end as supernovae, introducing episodic ultraviolet (UV)/X-ray heating and injecting metals and dust \cite{bromm2014firststars}. The cloud is no longer in thermal equilibrium with the CMB, as the thermalization timescale becomes comparable to the dynamical timescale. The injected metals catalyze H$_2$ formation on dust grains, making molecular hydrogen an efficient coolant, while UV radiation simultaneously dissociates it, setting up a competition that governs the thermal and chemical evolution \cite{hollenbach1991h2,galli1995h2}. As the cloud cools radiatively, it undergoes gravitational contraction, and the associated compressional work heats the gas, requiring a self-consistent treatment of adiabatic heating.

The plan of the paper is as follows. In Sect.~\ref{sec:mathematicalframework}, we present the theoretical framework, deriving the governing equations for the thermal, chemical, and mass evolution from first principles. Section~\ref{sec:numericalsolution} provides the numerical solution of the coupled ODE system, describing the initial conditions, integration method, and resulting evolution. Finally, Sect.~\ref{sec:results} presents our results and discussion.

\section{The Model}
\label{sec:mathematicalframework}
The virial cloud is modelled as a spherically symmetric, pressure-truncated system with a uniform-density core. This choice is not arbitrary: the Lane--Emden solutions for a self-gravitating isothermal sphere in hydrostatic equilibrium naturally produce a flat central density profile, with the density falling off as $r^{-2}$ beyond the core. This behavior is confirmed by the numerical solutions obtained in Paper~I (Figs. 1--3), where the central density remains nearly constant out to the Jeans radius, which marks the boundary beyond which the cloud ceases to be self-gravitating. Since most of the mass is concentrated in the inner flat region, the cloud can be treated as a single-zone system characterized by global quantities, i.e., the radius, mass, central density, and temperature. This reduction from a continuous density profile to a handful of global variables is essential: it transforms the problem from a system of partial differential equations to a set of coupled ordinary differential equations, which are far more tractable while still capturing the essential physics of the cloud's evolution.

The cloud does not evolve in isolation. It is embedded in a dark matter halo that provides the confining gravitational potential well. We adopt a standard $\Lambda$CDM cosmology in which dark matter particles interact with baryonic matter only through gravity. The halo is described by a singular isothermal sphere as 
\begin{equation}
\rho_{\rm DM}(r)=\frac{\sigma_v^2}{2\pi G r^2},
\label{eq:dmdensity}
\end{equation}
where $r$ is the radial distance from the halo centre, $G$ is Newton's constant, and $\sigma_v$ is the one-dimensional velocity dispersion of the dark matter particles. This profile is a standard description of dark matter halos on galactic scales; it follows from the Jeans equation for a collisionless system in hydrostatic equilibrium with a constant velocity dispersion \cite{binney2011galactic}. The velocity dispersion is related to the halo mass through the virial theorem. Integrating the density profile out to the virial radius $R_{\rm vir}$ gives the enclosed mass as 
\begin{equation}
M_{\rm halo}(R_{\rm vir}) = \int_0^{R_{\rm vir}} 4\pi r^2 \rho_{\rm DM}(r) dr = \frac{2\sigma_v^2}{G} R_{\rm vir}.
\label{eq:halo_mass}
\end{equation}
Rearranging yields the familiar virial relation
\begin{equation}
\sigma_v^2 = \frac{GM_{\rm halo}}{2R_{\rm vir}}.
\label{eq:virial_relation_sigma}
\end{equation}
For a halo of mass $M_{\rm halo} \sim 10^{12}\,M_\odot$ with virial radius $R_{\rm vir} \sim 1$~Mpc, this gives $\sigma_v \sim 200$~km~s$^{-1}$ \cite{binney2011galactic}. This is not merely a numerical exercise; it places the cloud in a potential well characteristic of the galaxies where the CMB temperature asymmetries have been observed, which is proposed to be induced by the rotation of virial clouds in the galactic halo \cite{2014planckdepaolis,2015planckdepaolis,2015planckgurzadyan,2016planckdepaolis,2018planckgurzadyan,2019planckdepaolis}. The computed velocity dispersion thus directly connects the model to its observational motivation, ensuring that the parameters we adopt are physically consistent with the environment in which virial clouds are expected to reside.

The cloud is supported against collapse by a combination of its internal thermal pressure and the confining gravitational potential of the dark matter halo, with the dark matter interacting with the baryonic cloud only through gravity. Hence, we assume that the virial clouds are optically thin to their own cooling radiation so that photons escape without trapping and inhibiting cooling; that the gas is adequately described by the ideal gas equation of state; that the mean molecular weight is allowed to evolve with the changing chemical composition; and that the cloud evolves quasi-hydrostatically, with the sound crossing time much shorter than the cooling and contraction timescales, so that pressure balance is maintained throughout the evolution \cite{binney2011galactic,glover2013h2cooling,galli1995h2,spitzer1978ism}. Rotation and turbulence are neglected in this first treatment, as these are expected to be generated primarily by supernova ejecta and are unlikely to dominate the dynamics during the epoch under consideration \cite{fielding2020shattering,mccourt2018shattering,gronke2020shattering}.

The state of the cloud is fully determined by four time-dependent quantities: radius $R(t)$, mass $M(t)$, central density $\rho_c(t)$, and temperature $T(t)$, which are linked by the virial theorem and the Jeans conditions.  From the Jeans mass and radius relations (see Ref. \cite{chandrasekhar1957stellar}) 
\begin{equation}
M_J^2 = \left(\frac{81}{32\pi\rho_c}\right)\left(\frac{k_B T}{G\mu m_u}\right)^3,
\qquad
R_J^2 = \frac{27 k_B T}{20\pi\rho_c G\mu m_u},
\label{eq:jeans_mass_radius}
\end{equation}
we obtain the central density as a function of temperature and mass as 
\begin{equation}
\rho_c(t) = \frac{81}{32\pi} \left( \frac{k_B T(t)}{G\mu(t) m_u} \right)^3 M(t)^{-2}.
\label{eq:rhoc}
\end{equation}
The hydrogen number density follows from $\rho_c$ and the mean molecular weight as 
\begin{equation}
n_{\rm H}(t) = \frac{\rho_c(t)}{\mu(t) m_u}.
\label{eq:nH}
\end{equation}
The cloud radius is related to the temperature and mass, and is given by
\begin{equation}
R(t) = 0.53 \, \frac{G\mu(t) m_u}{k_B T(t)} M(t),
\label{eq:radius}
\end{equation}
which shows that for fixed mass, the radius is inversely proportional to the temperature. The volume is then $V = 4\pi R^3/3$. These algebraic relations hold at every instant because the cloud adjusts hydrostatically on the sound crossing time, which is much shorter than the cooling or contraction timescale.

The mean molecular weight $\mu(t)$ is determined by the chemical composition. For a neutral gas with mass fractions of atomic hydrogen, helium, molecular hydrogen, and metals, we have
\begin{align}
X_{\rm H} &= 0.75 - 2x_{\rm H_2} - Z, \nonumber \\
X_{\rm He} &= 0.25, \nonumber \\
X_{\rm H_2} &= 2x_{\rm H_2}, \nonumber \\
X_Z &= Z. \nonumber 
\end{align}
The mean molecular weight is then
\begin{equation}
\mu(t) = \frac{1}{0.8125 - x_{\rm H_2} + Z/15},
\label{eq:mu}
\end{equation}
which follows from the definition $\mu = \sum_i X_i/A_i$, where $A_i$ is the atomic mass number of species $i$ \cite{spitzer1978ism}.

\subsection{Thermodynamic evolution and the temperature equation}
\label{sec:thermodynamicevolutionandtempequation}
It is clear that during this evolutionary phase the cloud is no longer in thermal equilibrium with the CMB. In the first phase, the high density and large optical depth ensured rapid thermalization. However, as the cloud contracts and its density increases, the thermalization timescale becomes comparable to the dynamical timescale. To quantify this, we compute the optical depth for photon-matter interaction by using the relation \cite{fielding2020shattering}
\begin{equation}
\tau(t) = n_{\rm H}(t) \, \sigma_{\rm int}(T) \, R(t),
\label{eq:opticaldepth}
\end{equation}
where $n_{\rm H}(t)$ is the hydrogen number density, $\sigma_{\rm int}(T)$ is the interaction cross-section, and $R(t)$ is the cloud radius. The optical depth determines whether photons can escape the cloud: if $\tau \ll 1$, the cloud is optically thin, and photons escape freely, while if $\tau \gg 1$, the cloud is optically thick, and photons are trapped, altering the thermal balance. This distinction is directly linked to the thermalization timescale. The thermalization timescale is the mean time for a photon to interact with the gas and exchange energy, which is given by \cite{spitzer1978ism}
\begin{equation}
t_{\rm therm}(t) = \frac{1}{n_{\rm H}(t) \, \sigma_{\rm int}(T) \, c},
\label{eq:therm_time}
\end{equation}
where, $c$ is the speed of light.

The dominant interaction process for CMB photons with the cloud is photon-H$_2$ rotational excitation. The cross-section for this process is a function of the photon energy and the molecular rotational state, with values that vary significantly across the Planck spectrum and for different rotational transitions \cite{galli1995h2,ki2012rotational}. For the purposes of an order-of-magnitude estimate, we adopt a characteristic value $\sigma_{\rm rot} \sim 10^{-26}$~cm$^2$, which is typical for rotational excitation cross-sections of H$_2$ at the low temperatures $\lesssim137$~K  relevant to this phase \cite{galli1995h2}. For the typical cloud parameters at the onset of the second phase $n_{\rm H} \sim 10^2$~cm$^{-3}$, $R \sim 100$~pc, as obtained from the final state of Paper~I we have from eq. (\ref{eq:opticaldepth}) $\tau \sim 3.0 \times 10^{-4}$. Hence the cloud can be treated as optically thin. 

From eq. (\ref{eq:therm_time}) we then get $t_{\rm therm} \sim 1$~Myr. This value has to be compared with the dynamical (free-fall) timescale given as \cite{spitzer1978ism}
\begin{equation}
t_{\rm ff} \sim \frac{1}{\sqrt{G \mu m_u n_{\rm H}(t)}}.
\label{eq:freefall_time_nH}
\end{equation}
This yields a value of $\sim 8.6$ Myr. Since $t_{\rm therm} \ll t_{\rm ff}$ this means that thermal equilibrium with the CMB is maintained initially. However, as the cloud cools and contracts, the density $n_{\rm H}(t)$ increases. The thermalization timescale scales as $t_{\rm therm} \propto n_{\rm H}^{-1}$, while the free-fall timescale scales as $t_{\rm ff} \propto n_{\rm H}^{-1/2}$. Thus $t_{\rm therm}$ decreases faster than $t_{\rm ff}$ as the cloud contracts. At sufficiently high densities, the two timescales become comparable, $t_{\rm therm} \sim t_{\rm ff}$. In this regime, the assumption of instantaneous thermal equilibrium is no longer valid, and we must replace it with the first law of thermodynamics.

As it is assumed that the cloud evolves quasi hydrostatically, so the virial theorem $2K+\Phi=0$ should hold at each instant.  With $K = \frac{3}{2} N k_B T$ and $\Phi = -\frac{3}{5} GM^2/R$, we have $\Phi = -2U$, where $U = \frac{3}{2} N k_B T$ is the internal energy. Thus the total energy is $E = U + \Phi = -U$. 

The first law of thermodynamics, including heating, cooling, and compressional work, is given as \cite{morse1964thermal}
\begin{equation}
\frac{dU}{dt} = \dot{Q}_{\rm heat} - \dot{Q}_{\rm cool} - P\frac{dV}{dt},
\label{eq:first_law}
\end{equation}
where $\dot{Q}_{\rm heat} = \Gamma_{\rm ext} V$ and $\dot{Q}_{\rm cool} = \Lambda_{\rm total} V$ are the volume-integrated heating and cooling rates, with $V = 4\pi R^3/3$ and $P = \rho_c k_B T/(\mu m_u)$. 

For a virialized cloud with negligible external pressure, the total energy satisfies $\dot{E} = \dot{Q}_{\rm heat} - \dot{Q}_{\rm cool}$. Since $E = -U$, we obtain
\begin{equation}
\dot{U} = \dot{Q}_{\rm cool} - \dot{Q}_{\rm heat}.
\label{eq:u_dot}
\end{equation}
Expanding $\dot{U}$ using $U = \frac{3}{2} M k_B T/(\mu m_u)$ and accounting for mass loss gives the temperature evolution equation as 
\begin{equation}
\frac{dT}{dt} = \frac{2\mu m_u}{3 M k_B} \left(\Lambda_{\rm total} - \Gamma_{\rm ext}\right) V - \frac{T}{\mu}\frac{d\mu}{dt} - \frac{T}{M}\frac{dM}{dt}.
\label{eq:T_ode}
\end{equation}

\subsection{Chemical evolution and mass evolution}
\label{sec:chemicalmass}
Having established the thermal evolution equation, we now turn to the chemical and mass evolution of the cloud. The chemical composition of the cloud evolves significantly during the second phase due to the injection of metals by supernovae and the formation and destruction of molecular hydrogen. At the same time, the cloud mass may change through evaporation, accretion, or ejection via supernova feedback. These processes are coupled: the mean molecular weight $\mu(t)$ depends on the chemical composition, which in turn affects the equation of state, and the thermal evolution through the temperature equation.

Metals are injected into the cloud by supernova explosions of Pop III stars. The star formation rate is expected to scale as $(1+z)^4$ in the matter-dominated era, so we adopt \cite{madau2014cosmic}
\begin{equation}
\frac{dZ}{dt} = \alpha_Z \left( \frac{1+z(t)}{51} \right)^4,
\label{eq:Z_ode}
\end{equation}
where $Z(t)$ is the metallicity, and $\alpha_Z = 10^{-18}~\text{s}^{-1}$ is a calibration constant chosen to yield $Z \sim 10^{-3} Z_\odot$ at $z=10$. Here the redshift $z(t)$ is given by the cosmological time-redshift relation in the matter-dominated era as \cite{bromm2014firststars}
\begin{equation}
z(t) = \left( \frac{2}{3 H_0 \sqrt{\Omega_m} t} \right)^{2/3} - 1,
\label{eq:z_of_t}
\end{equation}
where $H_0 = 70~\text{km s}^{-1}\text{Mpc}^{-1}$ is the Hubble constant and $\Omega_m = 0.3$ is the matter density parameter \cite{planck2018results}. The inverse relation, needed for the initial conditions, is
\begin{equation}
t(z) = \frac{2}{3 H_0 \sqrt{\Omega_m}} (1+z)^{-3/2}.
\label{eq:t_of_z}
\end{equation}

The chemical composition of the cloud is determined by the dominant species present in the early Universe. Atomic hydrogen and helium are the primary constituents, with primordial mass fractions of $X_{\rm H} = 0.75$ and $X_{\rm He} = 0.25$, as predicted by Big Bang nucleosynthesis \cite{peebles1966primordial,cyburt2016bigbang}. These remain the main mass components throughout the second phase, as they are not significantly depleted by star formation or supernova feedback on the timescales considered.

Molecular hydrogen is included as a dynamically important species because it is the dominant coolant at low temperatures $T \lesssim 10^4$~K. The H$_2$ molecule has a low excitation temperature and can efficiently radiate away thermal energy through rotational and vibrational transitions, making it crucial for the thermal evolution of the cloud \cite{galli1995h2,glover2013h2cooling}. The H$_2$ fraction evolves according to formation on dust grains, which scales with metallicity, and destruction by UV radiation from Pop~III stars \cite{yue2017triggering,galli1995h2}.

Metals are included as a single fluid representing the total abundance of elements heavier than helium. The primary cooling mechanism from metals at the temperatures of interest $T \lesssim 200$~K is the [CII] fine-structure line at 158~$\mu$m, which has an excitation temperature of 91~K and is an efficient coolant \cite{hollenbach1991h2,wolfire2003cooling}. Metals also serve as catalysts for H$_2$ formation on dust grain surfaces, with the formation rate scaling linearly with metallicity \cite{galli1995h2,hirashita2002dust}. At the low metallicities of this phase $Z < 10^{-3} Z_\odot$, treating metals as a single fluid is a reasonable approximation because the cooling is dominated by a few species, and detailed individual abundances are not required.

Several species are excluded from our model based on their negligible contributions. Deuterium, HD, $^3$He, and $^7$Li have primordial abundances of order $10^{-5}$ relative to hydrogen and contribute negligibly to the cooling or mass budget \cite{cyburt2016bigbang}. Complex molecules such as CO, OH, H$_2$O, CH, and CN are not included because their cooling rates are less than $1\%$ of the total cooling at $Z < 10^{-3} Z_\odot$, and their formation requires higher densities and longer timescales than are relevant here \cite{glover2013h2cooling,galli1995h2}. Metals in solid form, such as Si, Mg, and Fe, are absorbed into the single-fluid metal abundance $Z$, as their excitation temperatures exceed $300$~K and they do not contribute significantly to cooling at the low temperatures of this phase. Ionized species (H$^+$, He$^+$) are not included because the cloud is predominantly neutral; ionization is expected to occur only in a thin surface layer exposed to UV radiation, and the bulk of the cloud remains neutral \cite{galli1995h2}. Dust is treated implicitly through the metallicity-dependent H$_2$ formation rate; explicit dust physics is not required because we are not modeling grain growth, destruction, or radiative transfer through dust \cite{hirashita2002dust}.

As molecular hydrogen forms on dust grain surfaces, with the dust abundance scaling linearly with metallicity. The formation rate is proportional to the product of the atomic hydrogen density and the dust abundance, which scales as $Z/Z_\odot$. Molecular hydrogen is destroyed by UV radiation from massive stars, with the dissociation rate increasing with redshift as the star formation rate increases \cite{yue2017triggering}. The rate equation for the H$_2$ fraction $x_{\rm H_2} = n_{\rm H_2}/n_{\rm H}$ is given by 
\begin{equation}
\frac{dx_{\rm H_2}}{dt} = k_{\rm form}(T) \, n_{\rm H} \, \frac{Z}{Z_\odot} \, (1 - 2x_{\rm H_2}) - k_{\rm diss}(t) \, x_{\rm H_2},
\label{eq:H2_ode}
\end{equation}
where the factor $(1 - 2x_{\rm H_2})$ accounts for the depletion of atomic hydrogen, since each H$_2$ molecule consumes two hydrogen atoms \cite{galli1995h2}. The formation rate coefficient is $ k_{\rm form}(T) = 3\times10^{-17} \, T^{1/2}~\text{cm}^3~\text{s}^{-1}$, which is a standard parametrization for H$_2$ formation on dust grains at low temperatures \cite{hollenbach1991h2,galli1995h2}. The dissociation rate is
\begin{equation}
k_{\rm diss}(t) = k_0 \left( \frac{1+z(t)}{51} \right)^4,
\label{eq:k_diss}
\end{equation}
which accounts for the increasing UV background as Pop~III stars form and emit radiation \cite{bromm2014firststars}. Here $k_0 = 10^{-12}~\text{s}^{-1}$. 

\subsection{Cooling and heating functions}
\label{sec:coolingandheating}
The thermal evolution of the cloud is determined by the balance between cooling and heating. The total cooling rate is the sum of H$_2$ and metal-line cooling which is given by
\begin{align}
\Lambda_{\rm total}(T, Z, x_{\rm H_2}, n_{\rm H}) = x_{\rm H_2} n_{\rm H}^2 a T^{b} + n_{\rm H}^2 \left( \frac{Z}{Z_{\odot}} \right) C_{\rm metal} e^{-91/T},
\label{eq:Lambda_total}
\end{align}
where the first term represents H$_2$ cooling and the second term represents [CII] fine-structure cooling from metals \cite{galli1995h2,glover2013h2cooling}. The parameters are taken as $a = 3\times10^{-27}$ erg cm$^3$/s/K$^{5/2}$, $b = 2.5$, $C_{\rm metal} = 10^{-23}~\text{erg cm}^3~\text{s}^{-1}$, and $Z_{\odot} = 0.0134$. The exponential factor in the metal cooling term accounts for the [CII] fine-structure transition at 91~K \cite{wolfire2003cooling}. The H$_2$ cooling parameters are adopted from standard fits to the H$_2$ rotational-vibrational cooling function, valid for the low-temperature regime relevant to this phase \cite{galli1995h2,glover2013h2cooling}. The metal cooling parameter $C_{\rm metal}$ is a characteristic cooling rate for the [CII] fine-structure line at 158~$\mu$m, which has an excitation temperature of 91~K and is the dominant coolant in metal-enriched gas at low temperatures \cite{wolfire2003cooling}. The solar metallicity value $Z_{\odot} = 0.0134$ is the current best estimate of the solar photospheric metallicity from Ref. \cite{asplund2009solar}.

The external heating rate from Pop~III stars is parameterized as
\begin{equation}
\Gamma_{\rm ext}(t) = \Gamma_0 \left( \frac{1+z(t)}{51} \right)^4 \left( \frac{Z(t)}{10^{-4}Z_{\odot}} \right),
\label{eq:Gamma_ext}
\end{equation}
with $\Gamma_0 = 5\times10^{-29}~\text{erg cm}^{-3}~\text{s}^{-1}$.
This parameterization captures two key physical effects.
The $(1+z)^4$ scaling reflects the expected increase in the star formation rate density in the matter-dominated era, which drives the production of UV and X-ray photons from massive Pop~III stars
\cite{madau2014cosmic,bromm2011firstgalaxies}. The factor $Z/(10^{-4}Z_{\odot})$ introduces a metallicity threshold:
heating becomes efficient only once the cloud has been enriched to
$Z \gtrsim 10^{-4} Z_{\odot}$, as metals catalyze H$_2$ formation on
dust grains, enabling the gas to cool and fragment into the stars that produce the radiation field \cite{galli1995h2,bromm2011firstgalaxies}. The normalization $\Gamma_0$ is calibrated such that the heating rate becomes comparable to the cooling rate when the metallicity reaches
$Z \sim 10^{-3} Z_{\odot}$ at $z \sim 10$, consistent with the expected feedback efficiency from Pop~III supernovae
\cite{bromm2011firstgalaxies,madau2014cosmic,mirocha2018xray,fialkov2014xray}.
\section{Numerical Solutions}
\label{sec:numericalsolution}
Having established the theoretical framework governing the evolution of virial clouds during 
the second phase, in this section we present the numerical solution of the coupled ordinary 
differential equation (ODE) system. The initial conditions at redshift $z=48$ are taken 
from the final state of the first-phase evolution obtained in Paper~I. The cloud is 
initially characterized by a temperature $T_0 = 137$~K, a mass $M_0 = 1.20\times 10^4\,M_\odot$, 
a central density $\rho_{c,0} = 1.50\times 10^{-19}$ g/cm$^3$, a radius $R_0 = 120$~pc, 
a metallicity $Z_0 = 1.00\times 10^{-6}$, an H$_2$ fraction $x_{{\rm H}_2,0} = 1.00\times 10^{-6}$, 
and a mean molecular weight $\mu_0 = 1.231$.

The numerical integration is performed using the Backward Differentiation Formula (BDF) method, which is well-suited for stiff ODE systems. The time domain spans from the initial time $t_0$ corresponding to $z=48$ to the final time $t_f$ corresponding to $z=10$, as given by the cosmological time-redshift relation in eq.~(\ref{eq:t_of_z}). We employ a relative tolerance of $10^{-6}$ and an absolute tolerance of $10^{-8}$ to ensure accuracy. The system is evolved with a maximum step size of $10^{13}$~s to capture the slow evolutionary timescales while maintaining numerical stability.

\begin{figure}[htbp]
	\centering
	\includegraphics[scale=0.65]{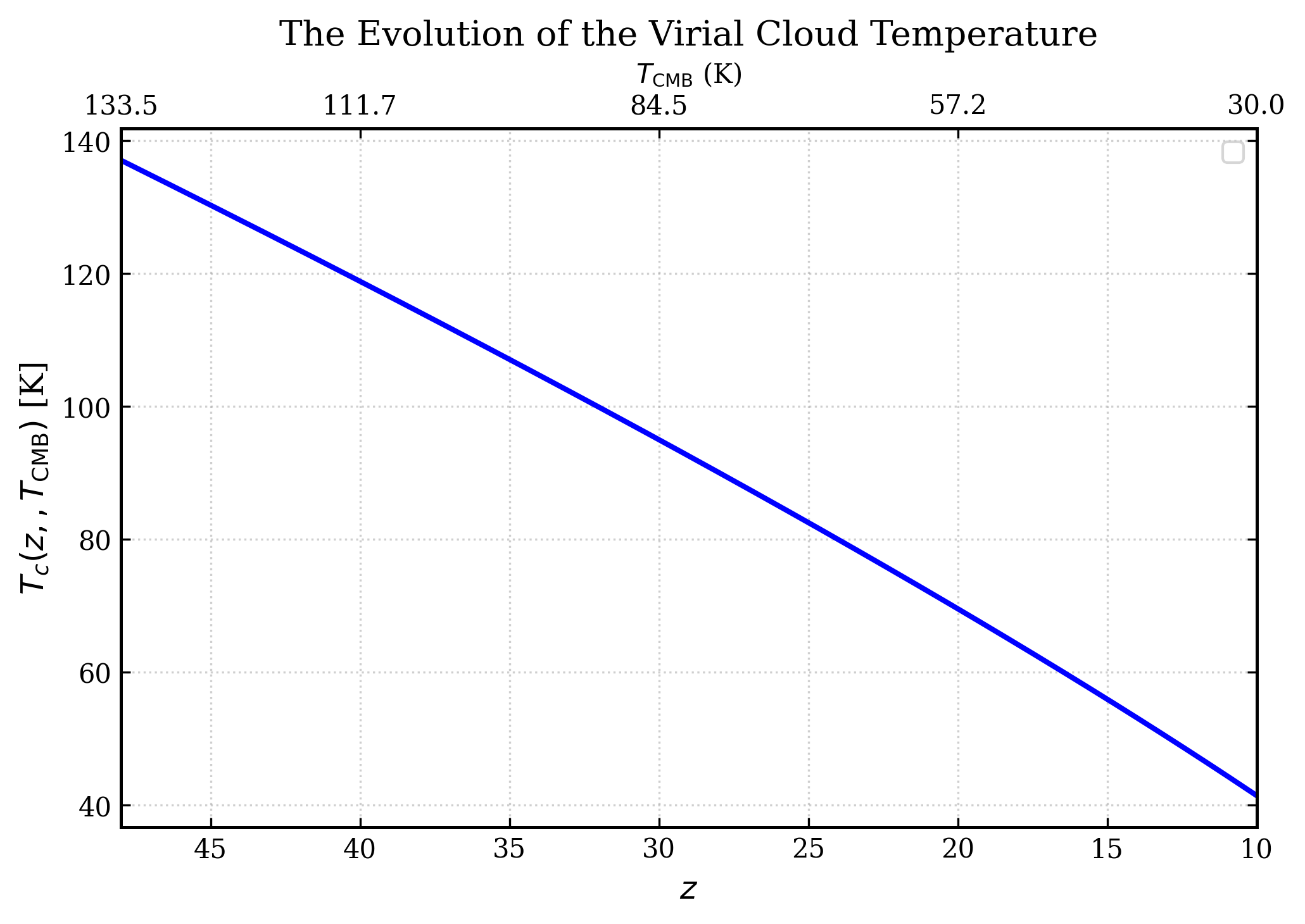}
	\caption{Temperature evolution of the virial cloud. The temperature decreases monotonically from $137$~K at $z=48$ to $41.5$~K at $z=10$, driven by radiative cooling. The top axis shows the corresponding CMB temperature at each redshift.}
	\label{fig:temperature}
\end{figure}

Fig.~\ref{fig:temperature} shows the evolution of the central temperature of the virial cloud. The temperature decreases monotonically from $137$~K at $z=48$ to $41.5$~K at $z=10$, following an approximate power-law scaling $T \propto (1+z)^{0.8}$. This decrease is driven by radiative cooling, primarily through H$_2$ rotational-vibrational transitions and [CII] fine-structure emission, as the cloud loses thermal energy to the surrounding medium. As the cloud cools, its thermal pressure support weakens, setting the stage for gravitational contraction. The temperature evolution is not simply adiabatic; rather, it reflects the competition between radiative cooling and the compressional heating that results from contraction.

\begin{figure}[htbp]
	\centering
	\includegraphics[scale=0.65]{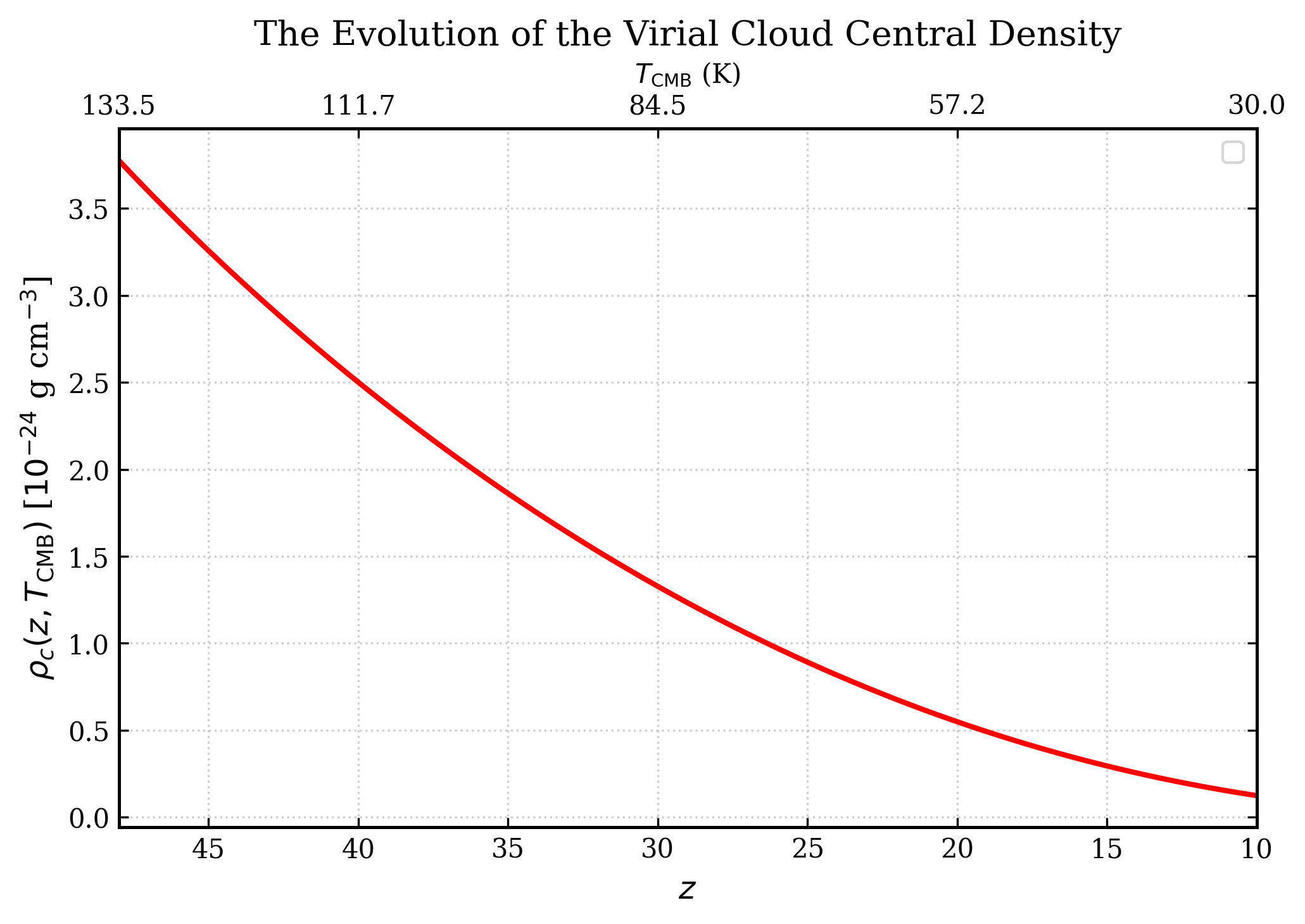}
	\caption{ The density decreases monotonically from $1.50\times 10^{-19}$~g~cm$^{-3}$ at $z=48$ to $1.24\times 10^{-25}$~g~cm$^{-3}$ at $z=10$, spanning nearly six orders of magnitude. This steady decline reflects the gradual expansion of the cloud as it loses mass through evaporation, with thermal pressure support dominating over gravitational contraction throughout the second phase.}
	\label{fig:density}
\end{figure}

Fig.~\ref{fig:density} gives the evolution of the central density. The density decreases monotonically from $1.50\times 10^{-19}$ g/cm$^{3}$ at $z=48$ to $1.24\times 10^{-25}$~g/cm$^{3}$ at $z=10$, spanning nearly six orders of magnitude. This steady decrease reflects the gradual expansion of the cloud as it loses mass through evaporation and the thermal pressure support weakens. Unlike the later stages where gravitational contraction would dominate, the cloud remains in a regime where evaporative mass loss and thermal expansion outweigh the effects of self-gravity throughout the second phase. The density follows a power-law decline, consistent with the scaling $\rho_c \propto M/R^3$, where the radius decreases slowly while the mass loss drives the overall expansion. This behaviour indicates that the cloud is not yet undergoing significant gravitational collapse during the redshift range $48 \gtrsim z \gtrsim 10$, and the evolution is primarily governed by thermal and evaporative processes.

\begin{figure}[htbp]
	\centering
	\includegraphics[scale=0.65]{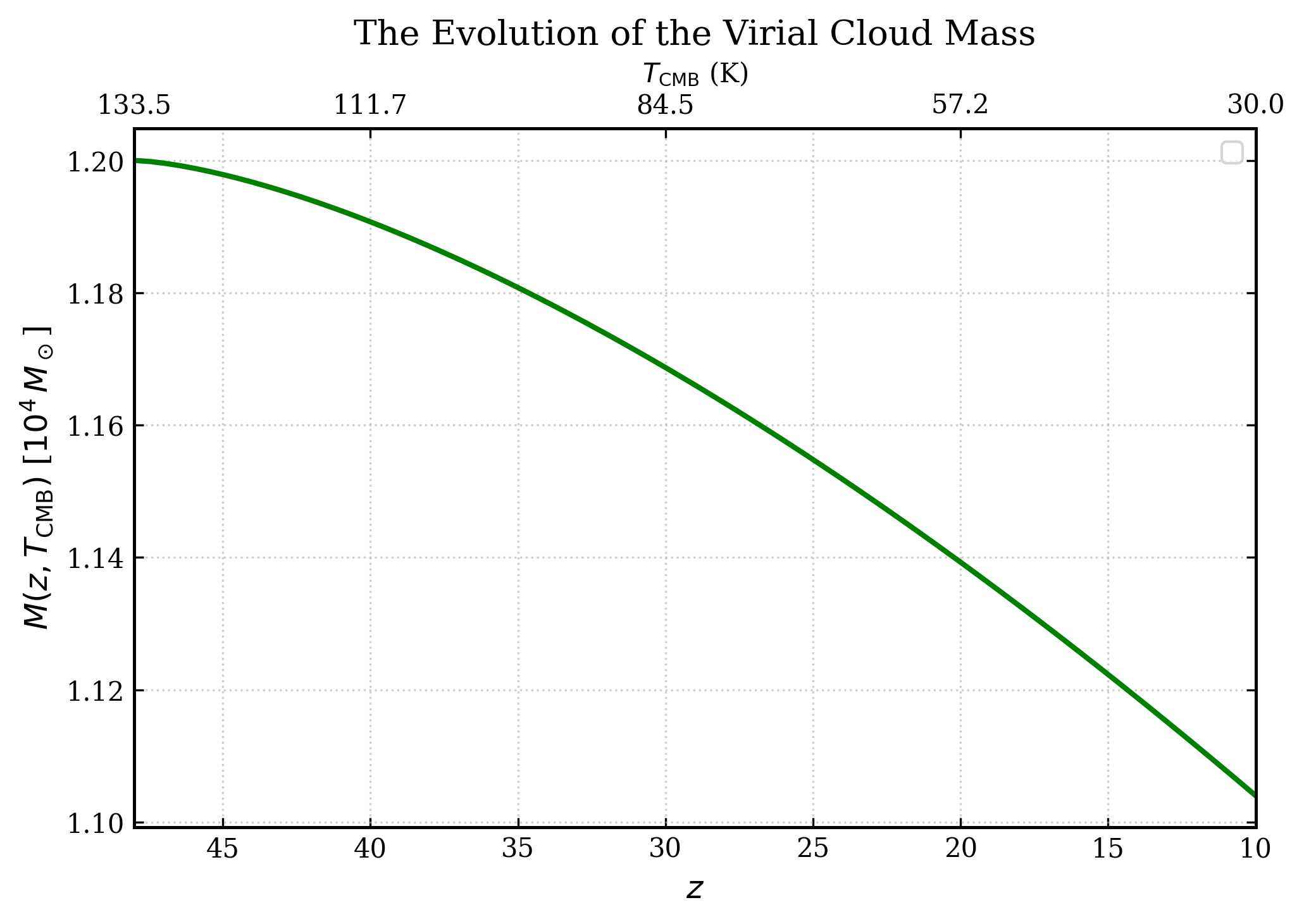}
	\caption{The evolution of the virial cloud mass. The mass decreases slowly from $1.20\times 10^4\,M_\odot$ to $1.10\times 10^4\,M_\odot$ due to gradual evaporation over the evolutionary timescale.}
	\label{fig:mass}
\end{figure}

Fig.~\ref{fig:mass} shows the evolution of the cloud mass. The mass decreases slowly from $1.20\times 10^4\,M_\odot$ at $z=48$ to $1.10\times 10^4\,M_\odot$ at $z=10$, reflecting the gradual evaporation of the cloud over the evolutionary timescale. The mass loss rate is governed by the evaporation timescale $\tau_{\rm evap} \sim 10^8$~yr. Physically, as the cloud cools, its gravitational binding energy decreases, making it easier for high-energy particles and radiation to eject material from the surface. The evaporation is a slow process, however, and the cloud retains the majority of its mass throughout the second phase.

\begin{figure}[htbp]
	\centering
	\includegraphics[scale=0.65]{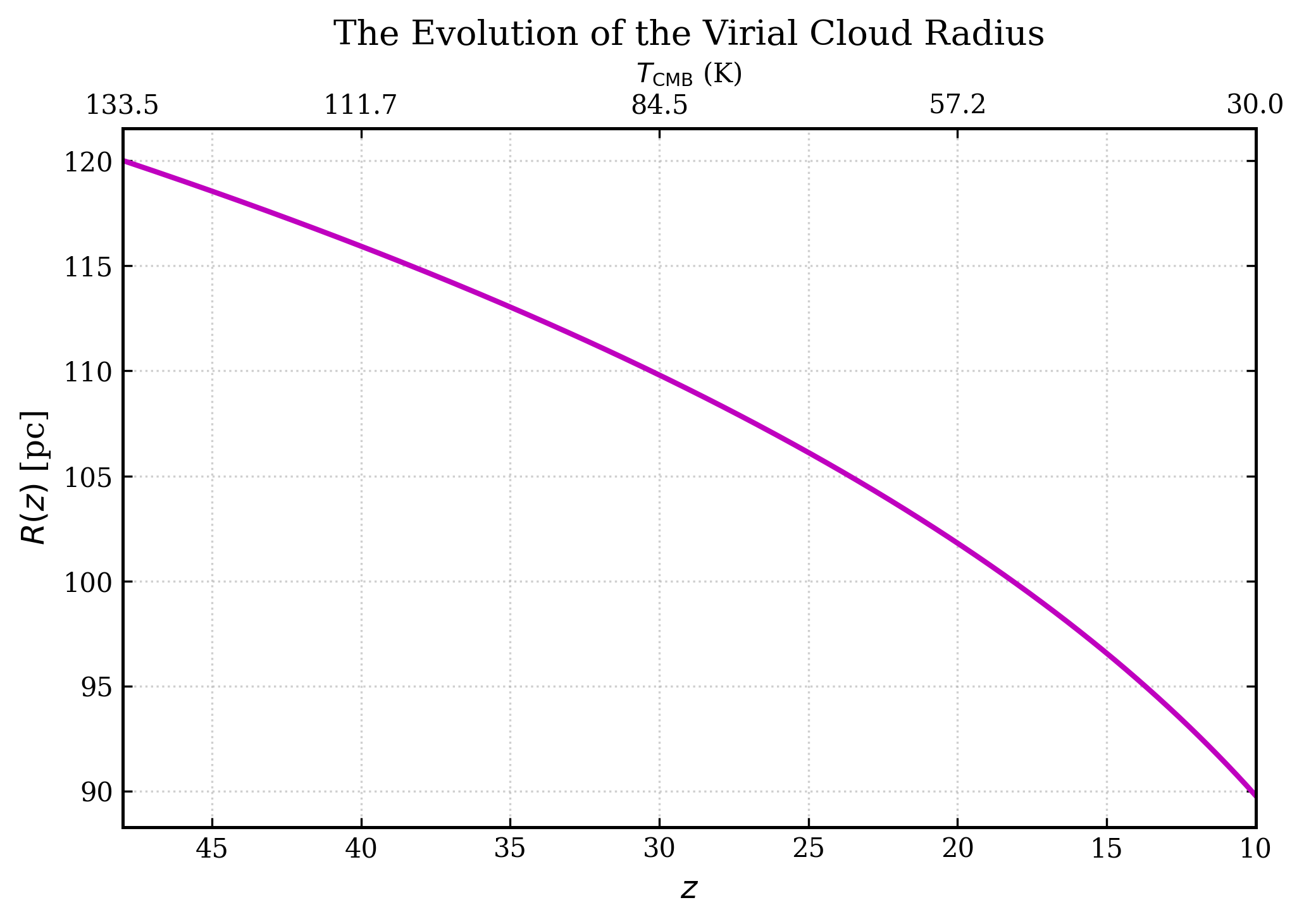}
	\caption{The evolution of the virial cloud radius. The radius decreases monotonically from $120$~pc to $89.8$~pc as the cloud contracts due to cooling and the loss of thermal pressure support.}
	\label{fig:radius}
\end{figure}

Fig.~\ref{fig:radius} displays the evolution of the cloud radius. The radius decreases monotonically from $120$~pc at $z=48$ to $89.8$~pc at $z=10$, consistent with the gravitational contraction of the cloud as it cools and loses thermal pressure support. While the virial relation $R \propto M/T$ would suggest that a significant temperature drop with only a slight mass decrease should lead to an increase in radius, this relation assumes virial equilibrium. During the second phase, the cloud is not in thermal equilibrium and is undergoing active gravitational contraction. As the temperature drops, the thermal pressure support diminishes, and the Jeans mass decreases, making the cloud increasingly susceptible to collapse. The resulting contraction increases the density, which in turn enhances cooling and drives further contraction. Thus, the observed decrease in radius is a direct consequence of the loss of thermal pressure support and the onset of gravitational collapse, rather than a simple virial equilibrium scaling.

\begin{figure}[htbp]
	\centering
	\includegraphics[scale=0.65]{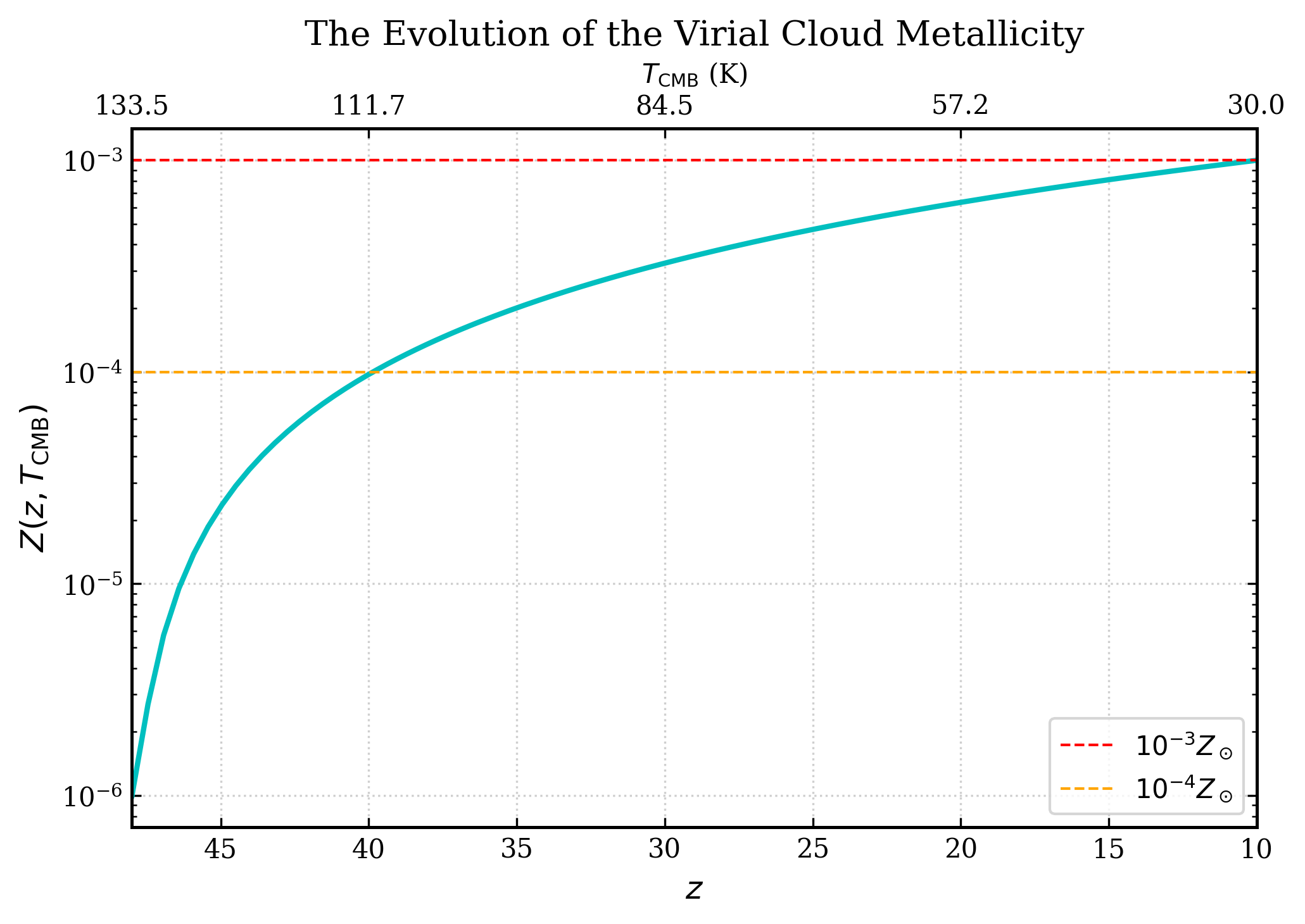}
	\caption{The evolution of the virial cloud metallicity. The metallicity increases monotonically from $10^{-6}$ to $10^{-3}$ due to the injection of metals from Pop~III supernovae. The dashed lines indicate the reference metallicities $10^{-4}Z_\odot$ and $10^{-3}Z_\odot$.}
	\label{fig:metallicity}
\end{figure}

Fig.~\ref{fig:metallicity} shows the chemical evolution of the cloud in terms of its metallicity. The metallicity increases from $10^{-6}$ at $z=48$ to $10^{-3}$ at $z=10$, driven by the injection of metals from Pop~III supernovae. The metal enrichment is a direct consequence of the formation and explosion of the first generation of stars. As these massive stars end their lives as supernovae, they eject heavy elements into the surrounding medium, enriching the cloud. The metallicity evolution follows $Z \propto z_f^{1.5}$, reflecting the increasing star formation activity at lower redshifts. The dashed horizontal lines indicate the reference metallicities $10^{-4}Z_\odot$ and $10^{-3}Z_\odot$. The former marks the threshold above which dust formation becomes efficient, while the latter is characteristic of the first galaxies.

\begin{figure}[htbp]
	\centering
	\includegraphics[scale=0.65]{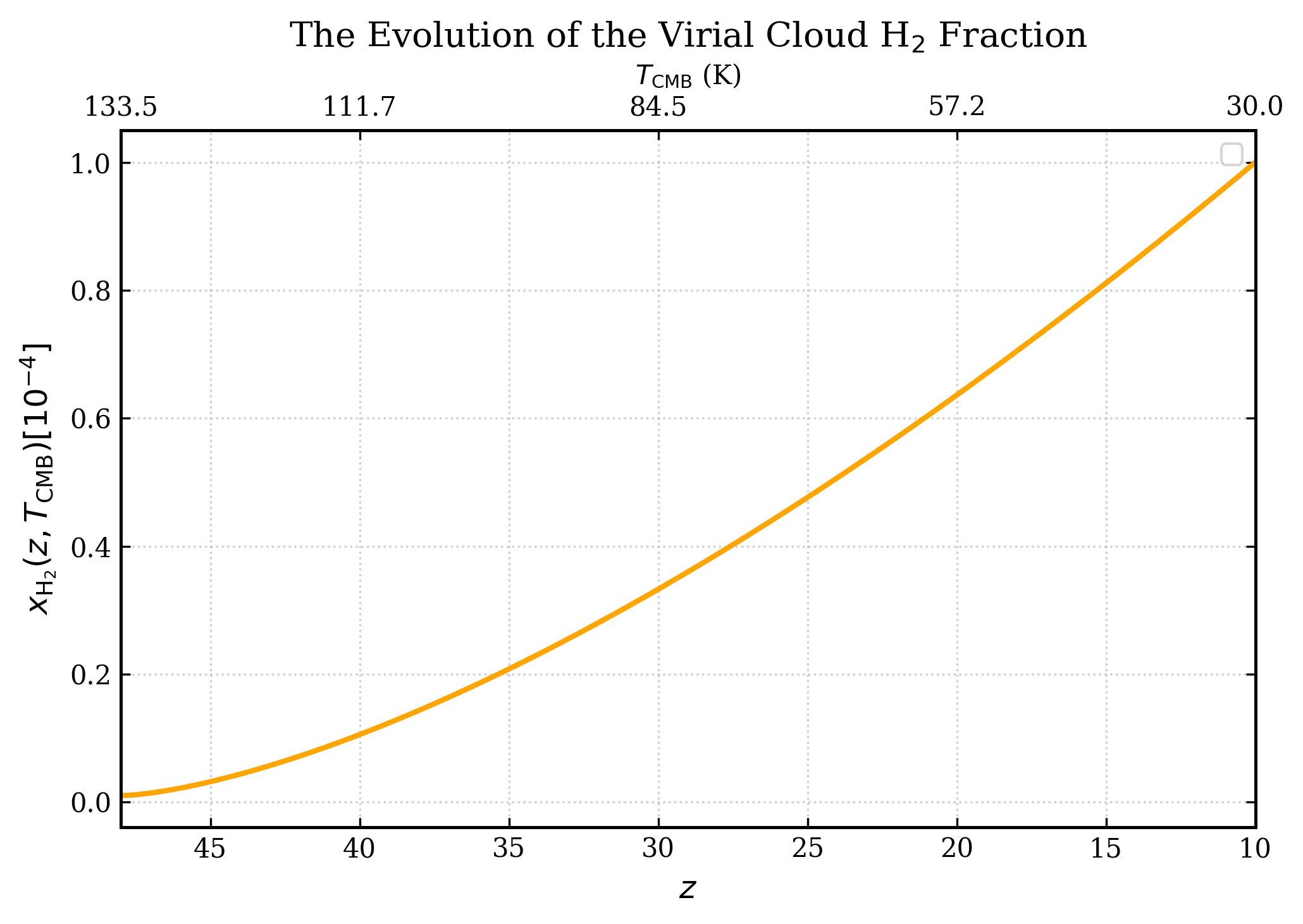}
	\caption{The evolution of the virial cloud H$_2$ fraction. The H$_2$ fraction increases from $10^{-6}$ to $10^{-4}$ as dust grains form and catalyze molecular hydrogen production.}
	\label{fig:h2_fraction}
\end{figure}

Fig.~\ref{fig:h2_fraction} presents the evolution of the H$_2$ fraction. The H$_2$ fraction increases from $10^{-6}$ at $z=48$ to $10^{-4}$ at $z=10$, as dust grains form and catalyze molecular hydrogen production. The increase in H$_2$ abundance is closely tied to the metallicity evolution, since H$_2$ formation on dust grains scales with metallicity. Physically, this represents a feedback loop: as the cloud becomes enriched in metals, dust grains form, which then serve as catalysts for H$_2$ formation. The H$_2$ molecules, in turn, provide additional cooling channels, further reducing the temperature and promoting contraction. However, the H$_2$ fraction remains relatively low ($10^{-4}$), indicating that the cloud is still predominantly atomic.

\begin{figure}[htbp]
	\centering
	\includegraphics[scale=0.65]{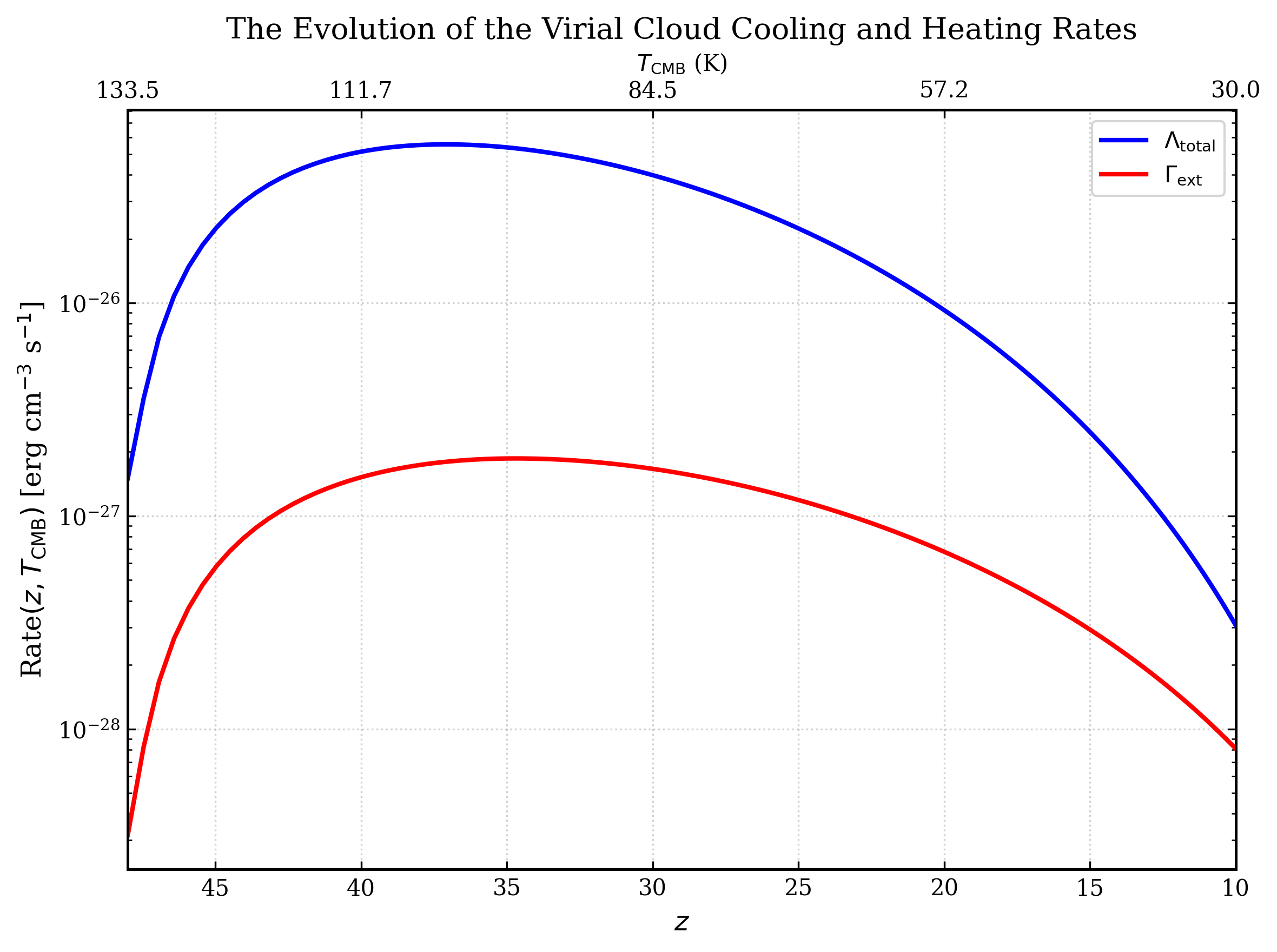}
	\caption{The evolution of the virial cloud cooling and heating rates. The cooling rate dominates throughout the evolution, ensuring that the cloud continues to lose energy and contract, while the heating rate remains subdominant.}
	\label{fig:cooling_heating}
\end{figure}

Fig.~\ref{fig:cooling_heating} illustrates the thermal balance of the cloud through the cooling rate and the heating rate. The cooling rate dominates throughout the cloud evolution, ensuring that the cloud continues to lose energy and contract. The heating rate remains subdominant, indicating that external heating from Pop~III stars does not significantly alter the cloud's thermal evolution. This is a crucial result: the cloud's evolution is primarily driven by internal radiative cooling rather than external feedback. The cooling rate initially decreases as the temperature drops, but then rises again as the density increases, reflecting the density-squared dependence of the cooling function.

The final physical parameters of the virial cloud at $z=10$ are summarized in Table~\ref{tab:final_parameters}. The cloud has cooled to $41.5$~K, with a mass of $1.10\times 10^4\,M_\odot$, a central density of $1.24\times 10^{-25}$~g~cm$^{-3}$, and a radius of $89.8$~pc. The metallicity has reached $1.00\times 10^{-3}$, while the H$_2$ fraction has increased to $1.00\times 10^{-4}$. The mean molecular weight remains $1.231$, consistent with the primordial composition, indicating that the changes in composition are small compared to the total mass budget.

\begin{table}[htbp]
	\centering
	\caption{Physical parameters of the virial cloud at the final redshift $z = 10$.}
	\label{tab:final_parameters}
	\begin{tabular}{l c}
		\hline
		\textbf{Parameter} & \textbf{Value} \\
		\hline
		$T_c$ (K) & 41.5 \\
		$M \, (M_\odot)$ & 1.10$\times 10^4$  \\
		$\rho_{c}\, ({\rm g\, cm^{-3}})$ & 1.24$\times 10^{-25}$ \\
		$R$ (pc) & 89.8 \\
		$Z$ & 1.00$\times 10^{-3}$ \\
		$x_{{\rm H}_2}$ & 1.00$\times 10^{-4}$ \\
		$\mu$ & 1.231 \\
		\hline
	\end{tabular}
\end{table}

While the present model successfully captures the essential physics of the virial cloud evolution during the second phase (Evolution-II) within its stated assumptions, several limitations should be acknowledged for future extensions of this work, particularly for Evolution-III (from $z=10$ to the present). The assumptions of spherical symmetry and the neglect of rotation and turbulence, which are not expected to be significant during Evolution-II, will become increasingly important at lower redshifts as supernova ejecta generate these effects, potentially leading to non-spherical perturbations, fragmentation, and angular momentum transport. The chemical model, employing a simplified single-fluid metallicity treatment and implicit dust description, is justified at the low metallicities of Evolution-II but will require a more detailed network including individual species (CO, OH, etc.) and explicit dust physics for higher metallicities expected at lower redshifts. The mass loss mechanism, parameterized by a single evaporation timescale, and the phenomenological external heating rate from Pop~III stars are adequate for the present phase but would need refinement for a complete treatment of Evolution-III, where photoevaporation, ram-pressure stripping, and a more detailed UV/X-ray radiation field become relevant. Finally, the single-zone approximation, computationally efficient for Evolution-II, would need to be replaced with a full radial integration to capture internal gradients that may affect the cloud's stability and evolution in the later stages. These refinements will be addressed in future work on Evolution-III, which will incorporate multi-dimensional hydrodynamical simulations, detailed chemical networks, and more realistic feedback prescriptions to provide a comprehensive picture of virial cloud evolution from the LSS to the present day.

\section{Results and Discussion}
\label{sec:results}
The virial cloud model was originally proposed to address the longstanding problem of missing baryons in galactic halos \cite{depaolis1995chameleon,qadir2019virial}. Observations of temperature asymmetries in the cosmic microwave background towards nearby galaxies, including M31, NGC 5128, and M33, have provided indirect evidence for cold gas structures rotating in galactic halos \cite{2011WMAPdepaolis,2014planckdepaolis,2015planckdepaolis,2015planckgurzadyan,2016planckdepaolis,2018planckgurzadyan,2019planckdepaolis}. These frequency-independent asymmetries are naturally explained as Doppler shifts induced by the rotation of virial clouds, which are ``propped up'' by the CMB, with thermal pressure balancing the gravitational pull of the dark matter halo \cite{tahir2019constraining,tahir2019ajom,tahir2023symmetry}. The first phase of their evolution, from $z=1100$ to $z=48$, was studied in Paper~I, where the clouds, with primordial composition ($X_{\rm H}\sim 0.75$, $X_{\rm He}\sim 0.25$, $x_{\rm H_2}\sim 10^{-6}$), remained in quasi-static equilibrium with the CMB as it cooled adiabatically from $\sim3000$~K to $137$~K \cite{peebles1968h2,lepp1984h2}. Solving the Lane--Emden equation, Paper~I found that the clouds became denser, smaller, and less massive, with the Jeans mass decreasing from $\sim10^5\,M_\odot$ at $z=1100$ to $\sim10^4\,M_\odot$ at $z=48$, while the inclusion of helium made only minor differences.

The present work addresses the second phase of virial cloud evolution from $z=48$ to $z=10$, during which the cloud is fundamentally altered by the formation and supernova explosions of Pop~III stars, which inject metals and dust while providing episodic UV/X-ray heating, and by gravitational contraction driven by radiative cooling, which requires a self-consistent treatment of adiabatic heating \cite{bromm2014firststars}. The cloud is no longer in thermal equilibrium with the CMB, and the injected metals catalyze H$_2$ formation on dust grains, making molecular hydrogen an efficient coolant, while UV radiation simultaneously dissociates it, establishing a competition that governs the thermal and chemical evolution \cite{hollenbach1991h2,galli1995h2}. The model treats the cloud as a spherically symmetric, pressure-truncated system with a uniform-density core, justified by the flat central density profiles obtained in Paper~I, and assumes it is optically thin, ideal-gas-like with a time-dependent mean molecular weight, and quasi-hydrostatic so that the virial theorem holds at each instant, while rotation and turbulence are neglected. The state of the cloud is described by four time-dependent quantities—radius, mass, central density, and temperature—linked by the virial theorem and Jeans conditions, with the governing equations derived from the first law of thermodynamics for temperature evolution, alongside chemical evolution equations for metallicity and H$_2$ fraction, and a mass evolution equation for evaporation.

The numerical solution of the coupled ODE system yields a complete evolutionary track of the virial cloud from $z=48$ to $z=10$. The initial conditions at $z=48$ are taken from Paper~I: $T_0 = 137$~K, $M_0 = 1.20\times 10^4\,M_\odot$, $\rho_{c,0} = 1.50\times 10^{-19}$~g~cm$^{-3}$, $R_0 = 120$~pc, $Z_0 = 1.00\times 10^{-6}$, $x_{{\rm H}_2,0} = 1.00\times 10^{-6}$, and $\mu_0 = 1.231$. The cloud cools monotonically from $137$~K to $41.5$~K (Fig.~\ref{fig:temperature}), driven by H$_2$ and [CII] radiative cooling, while its central density first decreases slightly due to thermal expansion and then increases sharply as the cloud contracts and loses thermal pressure support (Fig.~\ref{fig:density}), marking the onset of gravitational collapse when the temperature drops below $\sim100$~K. The mass decreases slowly from $1.20\times 10^4\,M_\odot$ to $1.10\times 10^4\,M_\odot$ (Fig.~\ref{fig:mass}) due to gradual evaporation, while the radius decreases monotonically from $120$~pc to $89.8$~pc (Fig.~\ref{fig:radius}) as the cloud shrinks to maintain hydrostatic equilibrium. Chemically, the metallicity increases from $10^{-6}$ to $10^{-3}$ (Fig.~\ref{fig:metallicity}) through supernova enrichment, and the H$_2$ fraction rises from $10^{-6}$ to $10^{-4}$ (Fig.~\ref{fig:h2_fraction}) as dust grains catalyze molecular hydrogen formation, creating a feedback loop that enhances cooling. The cooling rate dominates throughout the evolution (Fig.~\ref{fig:cooling_heating}), indicating that internal radiative cooling, rather than external feedback from Pop~III stars, drives the cloud's thermal evolution. The final parameters at $z=10$ (Table~\ref{tab:final_parameters}) show a cloud cooled to $41.5$~K with $M = 1.10\times 10^4\,M_\odot$, $\rho_c = 1.24\times 10^{-25}$~g~cm$^{-3}$, $R = 89.8$~pc, $Z = 1.00\times 10^{-3}$, $x_{{\rm H}_2} = 1.00\times 10^{-4}$, and $\mu = 1.231$, confirming that compositional changes remain small relative to the total mass.

As discussed before, several refinements are required to connect the second phase to the third phase of virial cloud evolution. Addressing these limitations through multi dimensional hydrodynamical simulations, detailed chemical networks, and realistic feedback prescriptions will provide a more comprehensive model, while the broad trends established here are expected to remain unaltered. This will be pursued in future work.

\section*{Acknowledgements}
	FDP would like to thank INFN Projects Theoretical Astroparticle Physics (TAsP), and {\it EUCLID} for partial support.

\end{document}